\documentclass[10pt,twocolumn,superscriptaddress,notitlepage,nofootinbib,nobibnotes,amssymb,amsmath]{revtex4-1}
\usepackage{slashed}
\usepackage{mathtools}
\usepackage{bm}

\usepackage{color}
\definecolor{purple}{rgb}{0.8,0,0.6}

\usepackage{tikz}

\newcommand{\beqn}{\begin{eqnarray}}
\newcommand{\eeqn}{\end{eqnarray}}
\newcommand{\eq}[1]{(\ref{#1})}

\newcommand{\cL}{{\cal L}}

\newcommand{\cl}{{\mathrm{cl}}}

\newcommand{\bs}{\boldsymbol}

\newcommand{\avr}[1]{{\left\langle #1 \right\rangle}}

\begin{document}
\bibliographystyle{naturemag}
\title{A Nernst current from the conformal anomaly in Dirac and Weyl  semimetals }

\author{M. N. Chernodub}
\affiliation{Laboratoire de Math\'ematiques et Physique Th\'eorique UMR 7350, Universit\'e de Tours, Tours 37200 France}
\affiliation{Laboratory of Physics of Living Matter, Far Eastern Federal University, Sukhanova 8, Vladivostok, 690950, Russia}

\author{Alberto Cortijo}
\affiliation{Instituto de Ciencia de Materiales de Madrid, CSIC, Cantoblanco; 28049 Madrid, Spain.}


\author{Mar\'ia A. H. Vozmediano}
\affiliation{Instituto de Ciencia de Materiales de Madrid, CSIC, Cantoblanco; 28049 Madrid, Spain.}


\begin{abstract}
We show that a conformal anomaly in Weyl/Dirac semimetals generates a bulk electric current perpendicular to a temperature gradient and the direction of a background magnetic field. The associated conductivity  of this novel  contribution to the Nernst effect is fixed by a beta function associated with the electric charge renormalization in the material.
\end{abstract}

\maketitle

Dirac and Weyl semimetals are three dimensional crystals whose low energy excitations are solutions of the massless Dirac  equation.  The recent experimental realization in a large family of materials \cite{Letal14a,Letal14b,Xu15,Lvetal15,Xuetal15} has provided unexpected  access to physical phenomena restricted so far to quite unreachable energy regions as the quark--gluon plasma \cite{KLetal12}. Quantum anomalies \cite{Shifman:1988zk} and anomaly--related transport \cite{Karl14} are at the center of interest of the actual research (an updated account is given in the reviews \cite{NaMa16,AMV17}). 
 
 After an intense activity around the experimental consequences of the axial anomaly \cite{KK13,XKetal15,Lietal15,ZXetal16} including evidences for the chiral magnetic effect \cite{LKetal16}, thermal transport is now probing gravitational anomalies \cite{CCetal14,Getal17}. The main link that opened the door to study gravitational effects in condensed matter systems is provided by the Luttinger theory of thermal transport coefficients \cite{Lut64,Stone:2012ud}.  He proposes  a  gravitational potential as the local source of energy flows and temperature fluctuations. The basic idea is that the effect of a temperature gradient that drives a system out of equilibrium can be compensated by a gravitational potential \cite{TE30}.  This advance completed the condensed matter description of thermo-electric-magnetic transport phenomena.  

A novel anomaly--induced transport phenomenon, the scale magnetic effect (SME) was described in a recent publication \cite{Chernodub:2016lbo}. Using massless QED as an example, it was shown that, in the background of an external magnetic field, the conformal anomaly \cite{CDJ77}  induces an electric current perpendicular to the magnetic field and to the gradient of the conformal factor. The coefficient was fixed by the beta function of the charge. In this work we show that a similar phenomenon will occur in Dirac and Weyl semimetals driven by a temperature gradient. The anomalous current:
\beqn
{\bs J} = \frac{e^2 v_F}{18 \pi^2 T \hbar} {\bs B}  \times {\bs \nabla} T\,,
\label{eq:C:SME:vF}
\eeqn
 provides a novel contribution similar to the Nernst effect occurring at zero chemical potential.
comes from the original SME with two important additions: First, the original result worked out in a conformally flat metric, has been extended to include smooth deformations from flat space what will allow to include material lattice deformations. The technical details of the derivation are done in the supplementary material. Second, we use the Luttinger construction to trade the conformal factor to a temperature gradient. Finally the Fermi velocity of the material $v_F$ will substitute the speed of light $c$ in the conductivity coefficient. In what follows we will detail these steps. 

The effective description of an interacting Dirac or Weyl semimetal around a single cone is given by the Lagrangian of massless QED in a flat Minkowski space-time. 
\beqn
\cL = - \frac{1}{4} F^{\mu\nu} F_{\mu\nu}  + {\bar \psi} i {\slashed D} \psi\,,
\label{eq:L:massless:QED}
\eeqn
where $\psi$ is the Dirac four spinor, $\bar \psi = \psi^\dagger \gamma^0$, ${\slashed D} = \gamma^\mu D_\mu$ with the covariant derivative $D_\mu = \partial_\mu - i e A_\mu$  and the Dirac matrices $\gamma^\mu$, and $F_{\mu\nu} = \partial_\mu A_\nu -  \partial_\nu A_\mu$ is the field strength tensor of the gauge field $A_\mu$. We notice that the electronic current: $J^\mu=\bar \psi(\gamma^0, v_F\gamma^i)\psi$ is anisotropic.  We will obviate this fact which does not play a role in this part.
The action  $S = \int d^4 x \, \cL$ of Eq.~\eq{eq:L:massless:QED} 
is invariant at a classical level under  a simultaneous rescaling of all coordinates and fields according to their canonical dimensions,
\beqn
x \to \lambda^{-1} x, 
\qquad 
A_\mu \to \lambda A_\mu, 
\qquad 
\psi \to \lambda^{3/2} \psi.
\label{eq:scale:transformation}
\eeqn

As a consequence of the scale invariance, the stress tensor of the model~\eq{eq:L:massless:QED},
\beqn
T^{\mu\nu} & = & - F^{\mu\alpha} F^\nu_{\  \alpha} + \frac{1}{4} \eta^{\mu\nu} F_{\alpha\beta} F^{\alpha\beta} 
\label{eq:Tmunu:QED}\\
& & + \frac{i}{2} {\bar \psi} \left(\gamma^\mu D^\nu + \gamma^\nu D^\mu \right) \psi - \eta^{\mu\nu} {\bar \psi} i {\slashed D} \psi\,,
\nonumber
\eeqn 
is  traceless, $(T^\mu_\mu)_{\cl} \equiv 0$. 
The scale invariance~\eq{eq:scale:transformation} is broken by quantum corrections which make the electric charge $e = e(\mu)$ dependent on the renormalization energy scale~$\mu$. As a result, in the background of a classical electromagnetic field $A_\mu$ the expectation value of the trace of the stress-energy tensor~\eq{eq:Tmunu:QED} becomes:~\cite{Shifman:1988zk}
\beqn
\avr{T^\alpha_{\ \alpha}(x)} = \frac{\beta(e)}{2 e}  F^{\mu\nu}(x) F_{\mu\nu}(x),
\label{eq:Tmunu:avr}
\eeqn
where $\beta(e)$ is the beta-function associated with the running coupling $e$:
$\beta(e) = \frac{{\mathrm d} e}{{\mathrm d} \ln \mu}\,.$
Hereafter we study quantum effects only in a classical electromagnetic background of the gauge fields $A_\mu \equiv A^{\mathrm{cl}}_\mu$.  

The conformal anomaly~\eq{eq:Tmunu:avr} leads to anomalous transport effects which most straightforwardly reveal themselves in a conformally flat space-time metric:
\beqn
g_{\mu\nu}(x) = e^{2 \tau(x)} \eta_{\mu\nu}\,,
\label{eq:g:munu}
\eeqn
where where $\tau(x)$ is a scalar conformal factor and $\eta_{\mu\nu}$ is the Minkowski metric tensor.

In a weakly curved ($|\tau| \ll 1$) gravitational background \eq{eq:g:munu} and in the presence of background magnetic field $\bs B$, the conformal (scale) anomaly~\eq{eq:Tmunu:avr} generates an anomalous electric current via the scale magnetic effect (SME):~\cite{Chernodub:2016lbo}
\beqn
{\bs J} & = & - \frac{2 \beta(e)}{e} {\bs \nabla} \tau(x) \times {\bs B}(x)\,.
\label{eq:SME}
\eeqn
In the presence of the electric field background $\bs E$ the conformal anomaly leads to the scale electric effect (SEE)
\beqn
{\bs J} = \sigma(x) {\bs E}(x)\,,
\label{eq:SEE}
\eeqn
which has the form of the Ohm law with the metric-dependent anomalous electric conductivity:~\cite{Chernodub:2016lbo}
\beqn
\sigma(t, {\bs x}) & = & - \frac{2 \beta(e)}{e} \frac{\partial \tau(t,{\bs x})}{\partial t}\,.
\label{eq:sigma:anomalous}
\eeqn
Both anomalous currents, \eq{eq:SME} and \eq{eq:SEE} can be described by the same relativistically covariant expression:
\beqn
J^\mu =  \frac{2 \beta(e)}{e} F^{\mu\nu} \partial_\nu \tau\,.
\label{eq:J:covariant}
\eeqn
The anomalous currents are generated in a quantum vacuum so that they emerge  at zero chemical potential and  in the absence of a classical current 
\beqn
J^\mu_{\mathrm{cl}} = - \partial_\nu F^{\mu\nu}\,,
\label{eq:J:cl}
\eeqn
in the space where the anomalous current  
is produced, $J^\mu_{\mathrm{cl}} (x) \equiv 0$.

Contrary to the axial anomaly, the scale anomaly is not exact in one loop. In particular, the beta function gets corrections at all orders in perturbation theory. The leading contribution to the current is defined by the one-loop QED beta function:
\beqn
\beta_{{\text{QED}}}^{\mathrm{1loop}} = \frac{e^3}{12 \pi^2}.
\label{eq:beta:QED}
\eeqn

A  generalization of Eq.~\eq{eq:J:covariant} to an arbitrary background metric is done in Appendix~\ref{appendix}. In our paper we consider the anomalous transport effects for gapless fermionic quasiparticles, realized in Weyl and Dirac semimetals, for which the conformal invariance is unbroken in the infrared region. For massive Dirac fermions the SME is strongly suppressed.\cite{Chernodub:2017bbd}

Having in mind condensed matter applications of our study, in the rest of the paper we will pay close attention only to the scale magnetic effect~\eq{eq:SME}. However, we would like to notice that its electric counterpart has certain interesting properties as well. For example, contrary to the usual Ohm conductivity, the anomalous conductivity~\eq{eq:sigma:anomalous} of the scale electric effect~\eq{eq:SEE} may take negative values. The negative vacuum conductivity, which may play a role in the Early Universe, has also been independent obtained in calculations for fermionic~\cite{Hayashinaka:2016qqn,Hayashinaka:2016qqnb} and bosonic~\cite{Kobayashi:2014zza} electrically charged particles in expanding de Sitter space via the Schwinger pair-production mechanism.

Now let us consider possible thermal effects which may play a role here. The basic idea is that the effect of a temperature gradient that drives a system out of equilibrium can be compensated by a gravitational potential $\Phi$:~\cite{Lut64,Stone:2012ud}
\beqn
\frac{1}{T} {\bs \nabla} T = - \frac{1}{c^2} {\bs \nabla}  \Phi\,,
\label{eq:Luttinger}
\eeqn
where $c$ is the speed of light. For weak gravitational fields the gravitational potential $\Phi$, to leading order, is related to the metric as follows:
\beqn
g_{00} = 1 + \frac{2 \Phi}{c^2}\,,
\label{eq:g00}
\eeqn
while other components of the metric tensor are unmodified. 

The electric current induced by the conformal effects is determined by Eq.~\eq{eq:J:SME} where the effective conformal factor is given by Eqs.~\eq{eq:v} and \eq{eq:g00}:
\beqn
\varphi(x) = - \frac{\Phi(x)}{3 c^2}.
\eeqn
In particular, for a time-independent gravitational potential $\Phi$ the scale electric effect is absent while the scale magnetic effect is given by Eq.~\eq{eq:SME} with the identification $\tau(x) \equiv \varphi(x)$. Thus, the current density given by the conformal anomaly is
\beqn
{\bs J} & = &  C_{\mathrm{conf}}\, {\bs B}  \times {\bs \nabla} T\,.
\label{eq:thermal:Nernst}
\eeqn
The conformal anomaly leads to a Nernst effect~\eq{eq:thermal:Nernst} with the coefficient described by the QED beta function~\eq{eq:beta:QED}:
\beqn
C_{\mathrm{conf}} = \frac{2\beta(e)}{3e} \equiv \frac{e^2 c}{18 \pi^2 T \hbar}\,,
\label{eq:C:SME}
\eeqn
where we have restored the powers of $\hbar$ and $c$. A similar strategy has been used in Ref.~\cite{Basar:2013qia} to derive a new correction to the Chiral Vortical Effect that arises in the presence of a temperature gradient.

To take into account the Fermi velocity we will now restore all $\hbar$ and $c$ in the fermionic Lagrangian in the SI system of units. The result~\eq{eq:thermal:Nernst} and \eq{eq:C:SME} corresponds to the Lagrangian 
\beqn
\cL =  {\bar \psi} \left[\gamma^0 \left( i \hbar \frac{\partial}{\partial t}  + e \phi \right) + c {\bs \gamma} \left( i \hbar {\bs \nabla }  - e {\bs A} \right)\right] \psi,\quad 
\label{eq:cL:c}
\eeqn
where we identified (in the SI units $A_t = \phi/c$):
\beqn
A^\mu = (\phi/c, {\bs A})\,,
\qquad
A_\mu = (\phi/c, - {\bs A})\,.
\label{eq:A}
\eeqn
The magnetic and electric fields are, respectively, as follows:
\beqn 
{\bs B} & = & {\bs \nabla } \times {\bs A}\,, \\
{\bs E} & = &  - {\bs \nabla } \phi - \frac{\partial {\bs A}}{\partial t}\,.
\eeqn

According to Eqs.~\eq{eq:thermal:Nernst} and \eq{eq:C:SME} the anomalous current corresponding to the Lagrangian~\eq{eq:cL:c} is:
\beqn
{\bs J} = \frac{e^2 c}{18 \pi^2 T \hbar} {\bs B}  \times {\bs \nabla} T\,.
\label{eq:C:SME:total}
\eeqn
Therefore we conclude that $c$ in the numerator of the current~\eq{eq:C:SME:total} is the $c$ which appears in the spatial derivative term of the Lagrangian~\eq{eq:cL:c}.
As mentioned before, the Lagrangian of Dirac and Weyl semimetals is

\beqn
\cL =  {\bar \psi} \left[i \gamma^0 \hbar \frac{\partial}{\partial t} + v_F {\bs \gamma} \left( i \hbar {\bs \nabla }  - e {\bs A} \right)\right] \psi,\quad 
\label{eq:cL:vF}
\eeqn
where we set $\phi=0$ as it does not affect our effect. Then the electromagnetic potential is
\beqn
A^\mu = (0, {\bs A})\,,
\label{eq:A:vF}
\eeqn
and the magnetic and electric fields are, respectively, as follows:
\beqn 
{\bs B} & = & {\bs \nabla } \times {\bs A}\,, \\
{\bs E} & = & - \frac{\partial {\bs A}}{\partial t}\,.
\eeqn

According to our considerations above the anomalous current corresponding to the Lagrangian~\eq{eq:cL:c} is:
\beqn
{\bs J} = \frac{e^2 v_F}{18 \pi^2 T \hbar} {\bs B}  \times {\bs \nabla} T\,.
\label{eq:C:SME:vF}
\eeqn

\begin{figure}[!thb]
\begin{center}
\includegraphics[scale=0.45,clip=true]{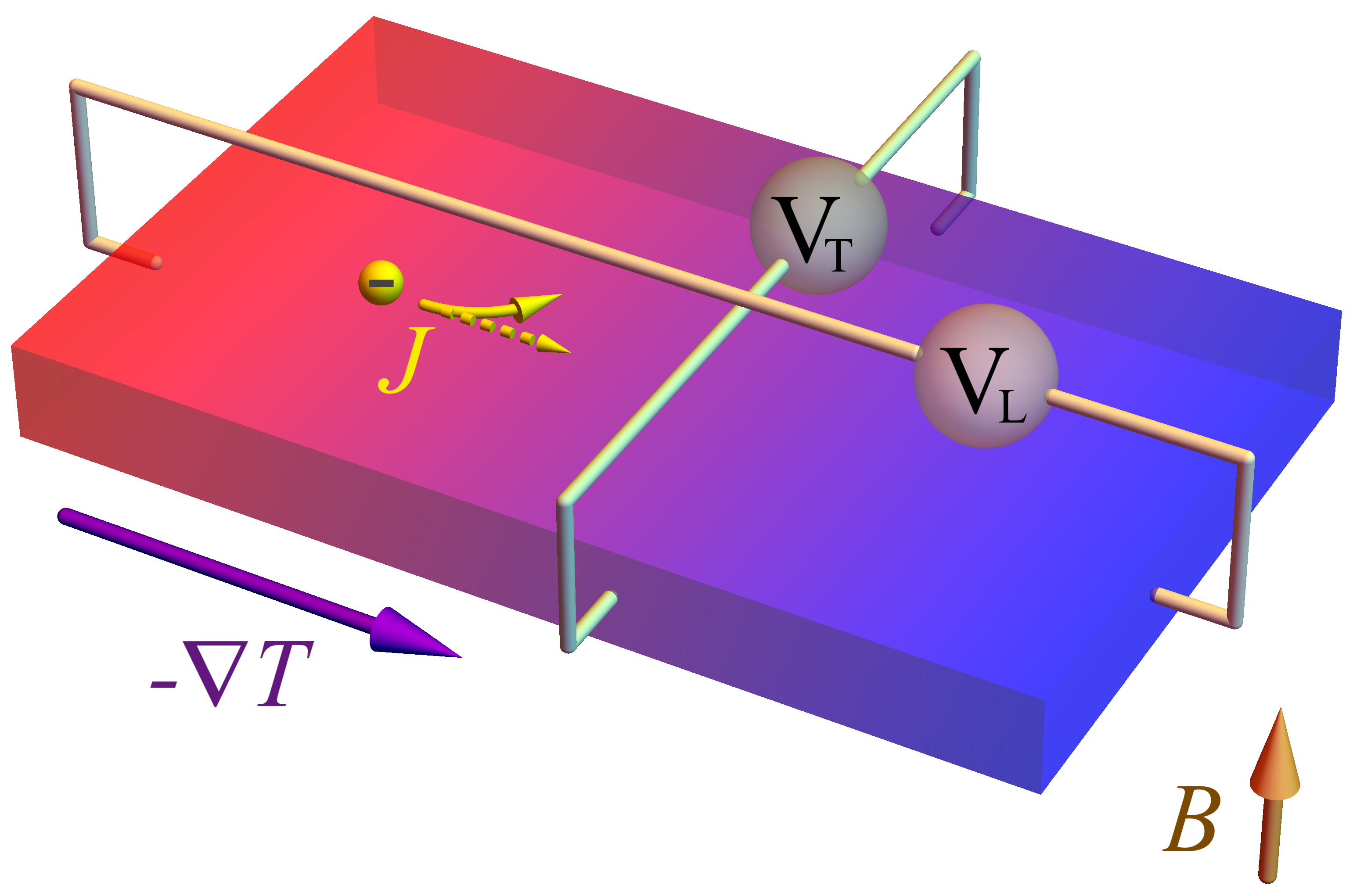}
\end{center}
\vskip -4mm 
\caption{(Color online) The setup of the Nernst-Ettingshausen effect in open circuit conditions. Voltage drops are induced by simultaneously applying an external magnetic field and a temperature gradient. Depending on if the measured voltage is perpendicular ($V_T$) or parallel to the gradient of $T$ ($V_L$), we speak about Nernst or Nernst-Ettingshausen effects (see details in the main text).}
\label{fig:scheme}
\end{figure}

To estimate the order of magnitude of the proposed effect, we have to remember that the Nernst effect is defined in open-circuit conditions, $\bm{J}=0$, thus appearing a voltage drop across the sample:
\beqn
J_{i}=\sigma_{ij}E_j+\mathcal{L}^{12}_{ir}(-\nabla_{r}T)=0,
\eeqn
(the notation $\mathcal{L}^{ab}_{ir}$ for transport coefficients is standard, and we have for instance, $\mathcal{L}^{11}_{ir}=\sigma_{ir}$. See, e.g. \cite{LLF14} for a modern reference). The induced electric field is, thus,
\beqn
E_{j}=\rho_{ji}\mathcal{L}^{12}_{ir}(-\nabla_{r}T),
\eeqn
where $\rho_{ji}=(\sigma^{-1})_{ji}$ is the resistivity tensor. For definitveness, let us choose the gradient of temperature to point, say, along the $x$ direction, $\nabla_1 T$, and the magnetic field $\bm{B}$ to point along $z$ as it is shown in Fig.~\ref{fig:scheme}. Then from Eq.~(48) the only component of the tensor $\mathcal{L}^{12}_{ir}$ is
\beqn
\mathcal{L}^{12}_{21}=\frac{e^2v_F B}{18\pi^2\hbar T}.
\eeqn

Under these conditions, two coefficients are usually defined. The Ettingshausen-Nernst coefficient is defined as
\beqn
S_{11}\equiv\frac{E_1}{B_3 \nabla_1 T}=\frac{\rho_{12}\mathcal{L}^{12}_{21}}{B},
\eeqn
and the Nernst coefficient,
\beqn
S_{12}\equiv\frac{E_2}{B_3 \nabla_1 T}=\frac{\rho_{22}\mathcal{L}^{12}_{21}}{B}.
\eeqn

In general, for three dimensional (isotropic) metals we have
\begin{subequations}
\beqn
\rho_{22}=\frac{\sigma_0}{\sigma^{2}_{0}+\sigma^2_{H}},
\eeqn
\beqn
\rho_{12}=\frac{\sigma_H}{\sigma^{2}_{0}+\sigma^2_{H}},
\eeqn
\end{subequations}
where $\sigma_0$ is the longitudinal conductivity and $\sigma_{H}$ is the transverse (Hall) conductivity. The longitudinal transport in undoped Weyl semimetals is strongly suppressed due to the absence of free carriers (transport coefficients are proportional to the chemical potential \cite{LLF14}), and the current is carried by counterpropagating electrons and holes\cite{HPV12}. However, at zero chemical potential Weyl semimetals have a finite topological anomalous Hall current:
\beqn
\bm{J}=\frac{e^2}{2\pi^2 \hbar}\bm{b}\times\bm{E},
\eeqn
where $\bm{b}$ is the separation between Weyl nodes. Choosing $\bm{b}$ to point along the $z$ direction, we have 
\beqn
\sigma_0\ll\sigma_{H}=\frac{e^2}{2\pi^2\hbar}|\bm{b}|, 
\eeqn
so $\rho_{22}\sim\frac{\sigma_0}{\sigma^2_{H}}$, and $\rho_{12}\sim\frac{1}{\sigma_H}$. 

The Ettingshausen-Nernst coefficient is then, approximately,
\beqn
S_{11}\equiv\frac{E_1}{B_3 \nabla_1 T}=\frac{\rho_{12}\mathcal{L}^{12}_{21}}{B}\sim\frac{v_F}{9|\bm{b}|T}.
\eeqn
The Nernst coefficient $S_{12}$ appears to be strongly suppressed due to $\sigma_{0}\ll \sigma_{H}$. For this reason, we propose to measure $S_{11}$. A small comment is in order here: it might be surprising that a transverse current as (\ref{eq:C:SME:vF}) leads to a longitudinal measurable quantity as it is $S_{11}$. The reason is that, due to the way the thermoelectric transports presented here are measured, the current in (\ref{eq:C:SME:vF}) is entangled to the resistivity tensor, which is dominated by the transverse Hall component, leading to a large coefficient $S_{11}$ compared with $S_{12}$.

For typical Fermi velocities in Weyl semimetals, $v_F\sim 10^5 m/s$, $T\sim 10 K$, and separation of Weyl nodes $|2\bm{b}|\sim0.3\AA^{-1}$, the Nernst coefficient divided by $T$ is of the order of $S_{11}/T\sim 0.6 \mu V/T K^{-2}$, which is of the same order of current Nernst measurements\cite{LLetal17}.

The importance of the Nernst and other thermo-magnetic effects for thermoelectric power generation, justifies the interest of the analysis of new sources  even if small in magnitude. The Nernst effect was explored in the early stages of novel Dirac materials  \cite{LLF14,SGT16,SMetal17} and some experimental results are already available in the literature \cite{LLetal17,WMetal17}. In most of the theoretical works the main ingredient are the magnetization of the materials or the Berry curvature acting as an effective magnetization in a semiclassical analysis. Our proposal is entirely new as a contribution to the Nernst current at zero chemical potential coming from the conformal anomaly.

\begin{acknowledgments}
We thank  A. G. Grushin and D. E. Kharzeev for useful conversations. We also thank K. Landsteiner for sharing his insight about this problem with us.
This work  has been supported by the PIC2016FR6/PICS07480, Spanish MECD
grant FIS2014-57432-P, the Comunidad de Madrid
MAD2D-CM Program (S2013/MIT-3007). A.C. acknowledges financial support through the MINECO/AEI/FEDER, UE Grant No. FIS2015-73454-JIN.
\end{acknowledgments}



\appendix
\section{Arbitrary gravitational fields}
\label{appendix}
\subsubsection{General gravitational background}
In the case of arbitrary -- i.e., not necessarily conformal and small -- metric $g_{\mu\nu}$ the effective anomalous action generated by one-loop quantum corrections has the following well known generally covariant form:~\cite{Riegert:1984kt,Mazur:2001aa,Mottola:2006ew,Armillis:2009pq}
\beqn 
S_{\mathrm{anom}}[g,A] & = & \frac{1}{8} \int d^4 x \sqrt{- g(x)}  \int d^4 y \sqrt{- g(y)} 
\label{eq:S:anom} \\
& & \hskip -21mm \cdot H(x) \Delta_4^{-1}(x,y)  \left[ 2 b C^2(y) + b' H(y) + 2 c F_{\mu\nu}(y) F^{\mu\nu}(y)\right]\!,
\nonumber
\eeqn
where the Weyl tensor squared 
\beqn
C^2 & = & C_{\mu\nu\alpha\beta} C^{\mu\nu\alpha\beta} \nonumber \\
& \equiv & R_{\mu\nu\alpha\beta} R^{\mu\nu\alpha\beta} - 2 R_{\mu\nu} R^{\mu\nu} + \frac{R^2}{3},
\label{eq:Weyl:tensor}
\eeqn
is expressed via the Riemann tensor $R_{\mu\nu\alpha\beta}$, the Ricci tensor $R_{\mu\nu} = R^{\alpha}_{\ \mu\alpha\nu}$ and the scalar curvature $R = R^\mu_{\ \mu}$. The linear combination 
\beqn
H = E - \frac{2}{3} \Box R\,,
\eeqn
involves the Euler (topological) density 
\beqn
E & = & {}^*R_{\mu\nu\alpha\beta} {}^*R^{\mu\nu\alpha\beta} \nonumber \\
& \equiv & R_{\mu\nu\alpha\beta} R^{\mu\nu\alpha\beta} - 4 R_{\mu\nu} R^{\mu\nu} + R^2,
\label{eq:Euler:density}
\eeqn
and the d'Alembertian differential operator $\Box \equiv \nabla^\mu \nabla_\mu$ of the scalar curvature $R$ expressed via the covariant derivative $\nabla_\mu$. Finally, 
\beqn
{}^*R_{\mu\nu\alpha\beta} = \frac{1}{2} \epsilon_{\mu\nu\mu'\nu'}R^{\mu'\nu'}_{\phantom{\mu'\nu}\alpha\beta}, 
\eeqn
is the (left) dual of the Riemann tensor $R_{\mu\nu\alpha\beta}$ and $g = \det g_{\mu\nu}$.

Due to the presence of the Green function~$\Delta_4^{-1}(x,y)$ of the fourth-order differential operator,
\beqn
\Delta_4 = \nabla_\mu \left( \nabla^\mu \nabla^\nu + 2 R^{\mu\nu} - \frac{2}{3} R g^{\mu\nu}\right) \nabla_\nu,
\label{eq:Delta:4}
\eeqn
the anomalous one-loop action~\eq{eq:S:anom} is a nonlocal function of the gauge field $A_\mu$ and metric $g_{\mu\nu}$. The nonlocality indicates that the scale anomaly is associated to an anomalous massless pole.

In massless QED~\eq{eq:L:massless:QED} the coefficients $b$, $b'$ and $c$ in the action~\eq{eq:S:anom} are, respectively, as follows:
\beqn
b = \frac{1}{320 \pi^2}, \qquad b' = - \frac{11}{5670 \pi^2},\qquad c = - \frac{e^2}{24 \pi^2}. \quad
\label{eq:bc}
\eeqn
The parameter $c$ is proportional to the one-loop QED beta function~\eq{eq:beta:QED}: $c = - \beta_{{\text{QED}}}^{\mathrm{1loop}}/(2e)$.

A variation of the action~\eq{eq:S:anom} with respect to metric gives us the correct expression for the one-loop trace anomaly:
\beqn
\avr{T^\mu_{\ \mu}} & \equiv & - \frac{2}{\sqrt{-g}} g_{\mu\nu} \frac{\delta S_{\mathrm{anom}}}{\delta g_{\mu\nu}} \nonumber \\
& = & - \frac{1}{4} \left[ b C^2 + b' \left( E - \frac{2}{3} \Box R \right) + c F_{\mu\nu} F^{\mu\nu}\right], \qquad
\label{eq:Tmunu:gravity}
\eeqn
while the classical (non-anomalous) part of the action does not contribute to the trace of the stress-energy tensor. Given the one-loop QED beta function~\eq{eq:beta:QED}, it is straightforward to show that the covariant trace~\eq{eq:Tmunu:gravity} reduces to Eq.~\eq{eq:Tmunu:avr} in a flat Minkowski space-time.

We are interested in the electromagnetic sector of the trace anomaly since the anomalous electric currents~\eq{eq:SME} and \eq{eq:SEE} are generated only in the presence of an external electromagnetic field $A_\mu$. 

The anomalous electric current is given by a variation of the anomalous action~\eq{eq:S:anom} with respect to the electromagnetic field $A_\mu$,
\beqn
J^{\mu}(x) & = & - \frac{1}{\sqrt{-g(x)}}\frac{\delta S_{\mathrm{anom}}}{\delta A_\mu(x)}\nonumber \\
& = & - \frac{1}{\sqrt{-g(x)}} \frac{\partial }{\partial x^\nu } \biggl[\sqrt{- g(x)}  \, F^{\mu\nu} (x) 
\label{eq:J:anom:exact} \\
& & \cdot \int d^4 y \sqrt{- g(y)} G(x,y) \Bigl(E(y) - \frac{2}{3} \Box R(y)\Bigr)\biggr]\,,\nonumber
\eeqn
where the Euler topological density is given in Eq.~\eq{eq:Euler:density} and the constant $c$ for QED with one species of fermion is given in Eq.~\eq{eq:bc}.

Equation~\eq{eq:J:anom:exact} is exact one-loop equation for anomalous electric current induced by conformal anomaly for arbitrary (not necessarily weak) gravitational field. Next we will discuss this current for weak gravitational fields and, in particular, for weak conformal fields.

\subsubsection{Weak gravitational background}

As we work with weak gravitational backgrounds, it is convenient to rewrite the electromagnetic part of the anomalous action~\eq{eq:S:anom},
\beqn
S^{(1)}_{\mathrm{anom}} & = & - \frac{c}{6} \int d^4 x \sqrt{- g(x)} \int d^4 y \sqrt{- g(y)} \nonumber \\
& & \cdot R^{(1)}(x) \, \square^{-1}_{x,y} \, F_{\alpha\beta} (y) F^{\alpha\beta} (y)\,,
\label{eq:nonlocal}
\eeqn
in terms a small perturbation ($|h_{\mu\nu}| \ll 1$) of the flat metric,
\beqn
g_{\mu\nu} = \eta_{\mu\nu} + h_{\mu\nu}\,.
\label{eq:g:h}
\eeqn
In Eq.~\eq{eq:nonlocal} the expression $\square^{-1}_{x,y}$ denotes a Green function of the flat-space d'Alembertian $\square \equiv \partial_\mu \partial^\mu$ and $R^{(1)}$ is the leading (linear in metric) double-derivative term of the Ricci scalar:
\beqn
R^{(1)} = \partial_\mu \partial_\nu h^{\mu\nu} - \eta_{\mu\nu} \square h^{\mu\nu}\,.
\eeqn
The indices are raised/lowered with the background metric tensor, $h^{\mu\nu} =  \eta^{\mu\alpha} \eta^{\nu\beta} h_{\alpha\beta}$. In the linearized gravity the inverse metric tensor is 
\beqn
g^{\mu\nu} = \eta^{\mu\nu} - h^{\mu\nu}\,,
\label{eq:ginv:munu}
\eeqn
[{\it cf.} Eq.~\eq{eq:g:h}], so that $g^{\mu\alpha} g_{\alpha\nu} = \delta^\mu_\nu + O(h^2)$.

In the conformally flat metric~\eq{eq:g:munu} with $|\tau| \ll 1$ one has $h_{\mu\nu} = 2 \tau \eta_{\mu\nu}$ so that $R^{(1)} = 6 \,\Box \tau$ and the leading contribution to the anomalous action~\eq{eq:nonlocal} reduces to 
\beqn
S^{(1),\mathrm{conf}}_{\mathrm{anom}} & = & - c \int d^4 x \int d^4 y [\square_x \tau(x)] 
\label{eq:nonlocal:tau} \\
& & \hskip 23mm \cdot\, \square^{-1}_{x,y} \, F_{\alpha\beta} (y) F^{\alpha\beta} (y)\,,
\nonumber
\eeqn
where subleading $O(\tau^2)$ terms are not shown. Integrating over the coordinate $y$ by parts in Eq.~\eq{eq:nonlocal:tau} and assuming that  the conformal perturbation of the metric $\tau$ vanishes in a spatial infinity, we obtain the local expression for the anomalous action in a weakly conformal background:
\beqn
S^{(1),\mathrm{conf}}_{\mathrm{anom}} = \frac{e^2}{24 \pi^2} \int d^4 x\, \tau(x) \, F_{\alpha\beta} (x) F^{\alpha\beta} (x)\,.
\label{eq:nonlocal:tau:local}
\eeqn
Hereafter we use the value of the parameter $c$ given in Eq.~\eq{eq:bc}.

A variation of the weak-field anomalous action~\eq{eq:nonlocal:tau:local} with respect to the electromagnetic field $A_\mu$ 
\beqn
J^{\mu}(x) = - \frac{1}{\sqrt{-g(x)}} \frac{\delta S^{(1)}_{\mathrm{anom}}}{\delta A_\mu(x)}\,,
\eeqn
provides us with Eq.~\eq{eq:J:covariant} which leads us to the scale magnetic~\eq{eq:SME} and scale electric~\eq{eq:SEE} effects.

In a general case of a weak gravitational fields~\eq{eq:g:h} the induced anomalous current density to leading order is as follows:
\beqn
J^{\mu}(x) = + \frac{e^2}{6 \pi^2} F^{\mu\nu}(x)\partial_\nu \varphi(x) 
\,,
\label{eq:J:SME}
\eeqn
where
\beqn
\varphi(x) & = & \frac{1}{6}  \int d^4 y \, \Box^{-1}_{x,y} \left[ \partial_\alpha \partial_\beta h^{\alpha\beta}(y) - \eta_{\alpha\beta} \Box h^{\alpha\beta}(y) \right]\nonumber \\
& \equiv & - \frac{1}{6}  \int d^4 y \, P_{\alpha\beta}(x,y) h^{\alpha\beta}(y)\,,
\label{eq:v}
\eeqn
where
\beqn
P_{\alpha\beta}(x,y)  = \eta_{\alpha\beta} \delta(x-y) - \Box^{-1}(x-y) \frac{\partial}{\partial y^\alpha} \frac{\partial}{\partial y^\beta} ,\quad
\label{eq:P}
\eeqn
is a transverse projector. In Eq.~\eq{eq:v} imposed the natural constraint that in the point $x$ the classical current~\eq{eq:J:cl} is absent, $J^\mu_{\mathrm{cl}} (x) = 0$. Again, for a weak conformally flat metric~\eq{eq:g:munu} we may identify the field~\eq{eq:v} with the conformal factor~\eq{eq:g:munu} of the metric, $\varphi(x) \equiv \tau(x)$ and we again come back to Eq.~\eq{eq:J:covariant} derived in Ref.~[\onlinecite{Chernodub:2016lbo}]. 

\end{document}